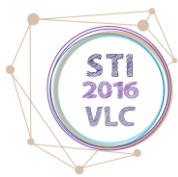

# Public scientists contributing to local literary fiction. An exploratory analysis[1]

Joaquín M. Azagra-Caro[*], Anabel Fernández-Mesa[**] and Nicolas Robinson-Garcia[*]

[*] jazagra@ingenio.upv.es; elrobin@ingenio.upv.es
INGENIO (CSIC-UPV), Universitat Politècnica de València, Valencia (Spain).
[**] Anabel.Fernandez@uv.es
Department of Management, Universitat de València, Valencia (Spain.

## INTRODUCTION

Public scientists (scientists only from now onwards), understood as a member of the teaching and/or research staff of a public university or a public research organization (including humanities and social sciences), benefit the academic community, industry and other social collectives through teaching and research. Most visible contributions are training skilled workforce, increasing the pool of knowledge and providing services to third parties. There are subtler contributions, like participation in unusual fora like diplomatic circles (Fähnrich, 2015) or informal exchange of ideas with social agents that convert scientists into part of the cultural core of cities (Florida, 2005). Active involvement of scientists in culture is part of the richness of developed societies. Some voices in current debates on the evaluation of societal impact and the role of universities towards social development are claiming a refocus from a socioeconomic perspective to also including sociocultural benefits from academia (Goddard, 2009).

In this paper we will focus in one facet of cultural engagement; writing literary fiction. We depart from considering that a full understanding of scientists' contributions should not only account for their research work but also other academic activities. Eminent examples of scientists that have written literary fiction are Nobel Prize Winners like José Echegaray, Toni Morrison, John Maxwell Coetzee or popular authors like J.R.R. Tolkien and Philip Roth. The objective of this research is to analyse the contribution of scientists to literary fiction. We define as such genre/commercial fiction, skipping the debate of whether genre/commercial literature is fiction with literary value. Literary fiction refers to the narrative forms of literature, i.e. novels, short stories, plays; thus leaving aside poetry or literary essay. Such contribution can be from local and non-local authors. We will narrow our general objective to local activities, due to the interest in the engagement of scientist on this geographic dimension.  Do local publishers include the literary work of scientists? Are works written by scientists more likely to be local than works not written by scientists?

Of course, scientists' literary fiction may not relate to their research. When there is a relation, literary fiction becomes a divulgation channel and, as such, a policy tool to increase knowledge diffusion and promote public understanding of science. This inquiry will also address a third research question: Do works written by local scientists present a higher degree of alignment between the contents of research and literary fiction?

---

[1] Nicolas Robinson-Garcia is currently supported by a Juan de la Cierva-Formación Fellowship from the Spanish Ministry of Economy and Competitiveness.

## DATA

Here we present preliminary result of an exploratory pilot study focused on a local publisher from the city of Valencia, Ediciones Contrabando and books published in 2015. By the date of the conference, we plan to expand to all publishers located in Valencia publishing literary works as defined above. Ediciones Contrabando was an interesting pilot study because it is a young company (born in 2012) and we have personal contacts with the chief editor, which is useful to explore in-depth issues that may arise when performing a larger scale analysis as well as work with a sample from which we can gain full understanding. Data was extracted from the Spanish national database of books published in Spain[2]. This portal includes a database of Spanish publishers and a database of books. Ediciones Contrabando published in 2015 a total of 17 books. By manually checking online information about the books, we configured a valid sample of 9 fiction works (7 were non-fiction, 1 was not identifiable and 1 was a second edition of a book published the same year, which we excluded).

A well reported issue when working with monographs is the lack of address information of the authors (Gorraiz, Purnell & Glänzel, 2013), this makes it problematic when trying to identify the institutions behind such works or the background of the author. Here, the identification of scientists was done manually by checking in the Internet and searching for information on the author's affiliation.

## METHOD

In order to measure the contribution or scientists to literary fiction, our proposal is to identify which proportion P of edited books of literary fiction is authored by scientists. We will answer our first research question by estimating the model:

$$P_{ijkl} = f(L_j, Z_{ijkl})$$

Here, P is the probability that the edition i of book j by author k, published by company k, has an academic author. L represents whether the author is from the same city of the publishing company. Z is a vector of control characteristics.

For academic authors, we will measure the degree of alignment D between their research and their fiction work. We will consider three possible degrees: 0 (none), 1 (formal alignment, i.e. the action occurs in a scientific setting), 2 (content alignment, i.e. the plot has to do with the scientist's research lines). For further analysis, we will estimate the model:

$$D_{ijkl} = f(L_j, Z_{ijkl})$$

## PRELIMINARY RESULTS

Ediciones Contrabando published 9 fiction books in 2015. 3 had scientific authors, i.e. 33%. This is a first quantification of the contribution of public scientists to literary fiction. At first sight, probably an overestimation that further research and enlargement of the dataset will nuance, but also a manifestation the phenomenon is present.

---

[2] http://www.mecd.gob.es/cultura-mecd/areas-cultura/libro/bases-de-datos-del-isbn.html

Local authors were 7, i.e. 78%. This is coherent with the intuition that young, small editorial companies rely on geographic proximity to nurture their portfolios. The 3 scientific authors were local, which implies that the localisation between editor and authors increases the probability of finding scientific authors. By the day of the conference, we will test this via econometric models with a larger number of observations.

The degree of alignment between the research lines of the scientific authors and their fiction is null in the three cases. Hence, in this pilot study, we would observe that scientific authors do not diffuse academic knowledge through their literary works, and that co-location with the editor does not play a role.

## DISCUSSION AND FURTHER RESEARCH

As members of publicly-funded institutions, scientists have a central role on the creation of new knowledge in order to facilitate social progress. Lately, a key concern for policy makers is to learn about the societal contribution of scientists for social development. In this sense, contributions are normally understood in terms of research activity and output, and not in terms of the implications of having an academic population embedded in a local environment. Here we build on the creative class of Florida (2005) where it is suggested that the inclusion of a certain group of individuals enhances local development and enrichment. We propose to analyse their role in literary fiction, specifically we are interested in analysing the level of alignment between their work as academics and their ouvre, and see if they use it as a tool for facilitating the public understanding of science. As 'civic' universities where researchers are viewed as workers for the public good, activities such as literary publishing can serve as powerful tools for popularizing science (Goddard, 2009).

Here we propose a first approach based on local publishers in a given city. This way we can control variables such as the propensity of scientific authors to publish with local companies, the relation between having research institutions and a larger number of literary scientists, etc. However, we still need to overcome important limitations. Here we show a very small sample of books from a small publishing company. The difficulty to automatically identify the affiliation of authors remains a considerable shortcoming for working with large datasets which could provide us a better understanding and insight of the role of scientists in local literary publishing. Finally, we must point out other issues that will be solved in the future such as the identification of genre and other classical problems when working with monographs (Torres-Salinas et al., 2014) such as translations, new editions, edited books of short stories, etc.